\documentclass[slac_one]{revtex4}
\usepackage{graphicx}
\usepackage{fancyhdr}
\begin{document}
\newcommand{\Bo}{B^{0}}
\newcommand{\Bp}{B^{+}}
\newcommand{\Bm}{B^{-}}
\newcommand{\Dp}{D^{+}}
\newcommand{\Dm}{D^{-}}
\newcommand{\Do}{D^{0}}
\newcommand{\Dob}{\overline{D}^{0}}
\newcommand{\Dst}{D^{*}}
\newcommand{\Dsto}{D^{*0}}
\newcommand{\Dstb}{\overline{D}^{*}}
\newcommand{\Dstp}{D^{*+}}
\newcommand{\Dstm}{D^{*-}}
\newcommand{\DDst}{DD^{*}}
\newcommand{\Kp}{K^{+}}
\newcommand{\Km}{K^{-}}
\newcommand{\Ks}{K_{s}^0}
\newcommand{\pio}{\pi^{0}}
\newcommand{\pip}{\pi^{+}}
\newcommand{\pim}{\pi^{-}}

\newcommand{\ddgamma}{\Dsto(\Do\gamma)\Dob}
\newcommand{\ddpio}{\Dsto(\Do\pio)\Dob}
\newcommand{\mbc}{M_{\mathrm{bc}}}
\newcommand{\pqr}{E_9/E_{25}}
\title{\boldmath{$CP$} violation in \boldmath{$B$} decays to charm and charmonium at Belle} %% Paper title goes here


\author{K. Vervink}
\affiliation{Ecole Polytechnique F\'ed\'erale de Lausanne (EPFL), Switzerland}

\begin{abstract}
We present the study of $CP$ violation in charm and charmonium decays, using
a data sample corresponding to $657 \times 10^6$ $B\overline{B}$ events 
collected with the Belle detector at the $\Upsilon(4S)$ resonance at the KEKB
asymmetric-energy $e^+e^-$ collider.  We report measurements of the
polarization fraction and
time-dependent $CP$-violation parameters of the decay $B^0\to D^{*+}D^{*-}$
and of the branching fraction and charge asymmetry in the decay of the
Cabibbo- and color-suppressed process $B^{\pm} \to \psi(2S) \pi^{\pm}$.
\end{abstract}

\maketitle

\thispagestyle{fancy}


\section{INTRODUCTION} % Section title should be in all capitals.
The study of exclusive $B$ meson decays to charm and charmonium has
played an important role in exploring $CP$
violation~\cite{bib:charmonium, bib:charm, bib:miyake}.  Amongst them the
Cabibbo-suppressed decays have an increased sensitivity to New Physics
effects and can be studied at the $B$ factories which, due to their
high integrated luminosities, overcome the suppression factor.
The $B^0 \to D^{*+} D^{*-}$ and $B^{-} \to \psi(2S) \pi^{-}$ decay proceed
primarily via a $b \to c\bar{c}d$ tree
diagram while penguin contributions are expected to be small in the
Standard Model (SM).   A large deviation of the measured $CP$
parameters from the SM prediction can be a hint of New Physics.

\section{DATASET AND EVENT RECONSTRUCTION}
Both analyses are based on a data sample containing $657$ million $B \overline{B}$ pairs,
collected with the Belle detector~\cite{bib:det} at the KEKB asymmetric-energy  $e^+
e^-$ collider~\cite{bib:kekb} operating at the $\Upsilon(4S)$
resonance.  The $\Upsilon(4S)$ meson is produced with a Lorentz
boost $\beta \gamma = 0.425$ along the $z$ axis, opposite to
the positron beam direction, and decays mainly into a $B^0
\overline{B}^0$ or a $B^+ B^-$ pair.

The reconstructed $B$ candidates are discriminated from
background using the energy difference 
$\Delta E \equiv E_{B}^{{\rm CM}} - E_{{\rm beam}}^{{\rm CM}}$ and the beam-constrained mass $M_{{\rm bc}} \equiv \sqrt{(E_{{\rm beam}}^{{\rm CM}})^2 -
  (p_{B}^{{\rm CM}})^2}$,  where $E_{{\rm beam}}^{{\rm CM}}$ is the beam
energy in the center-of-mass (CM) system and $E_{B}^{{\rm CM}}$ and
$p_{B}^{{\rm CM}}$ are 
the CM energy and momentum of the $B$ candidate.

\section{\boldmath{$B^- \to \psi(2S) \pi^-$}}

\subsection{Theoretical Motivation}
The main $B^{-} \to \psi(2S) \pi^{-}$ diagram is not only
Cabbibo-suppressed but also color-suppressed.  
A measurement of its
branching fraction, unknown so far, 
is shown in this paper.  Assuming tree dominance and
factorization, the branching fraction $\mathcal{B}(B^- \to \psi(2S)
\pi^-$) is expected to be about $5\%$ of that of the Cabibbo-favored
mode $B^- \to \psi(2S) K^-$~\cite{bib:neubert}.  Furthermore under these
assumptions, $CP$ violation should be negligibly small.  However if
penguin contributions or new physics effects are present, a non-zero
charge asymmetry can occur.

\subsection{Results}
%%%%%%%%%%%%%%%%%%%%%%%%%%%%%%%%%%%%%%%%%%%%%%%%%%%%%%%%%%%%%%%%%%%%%%%%%%%%%%%%%%
\begin{figure}
\includegraphics[width= 0.49\textwidth]{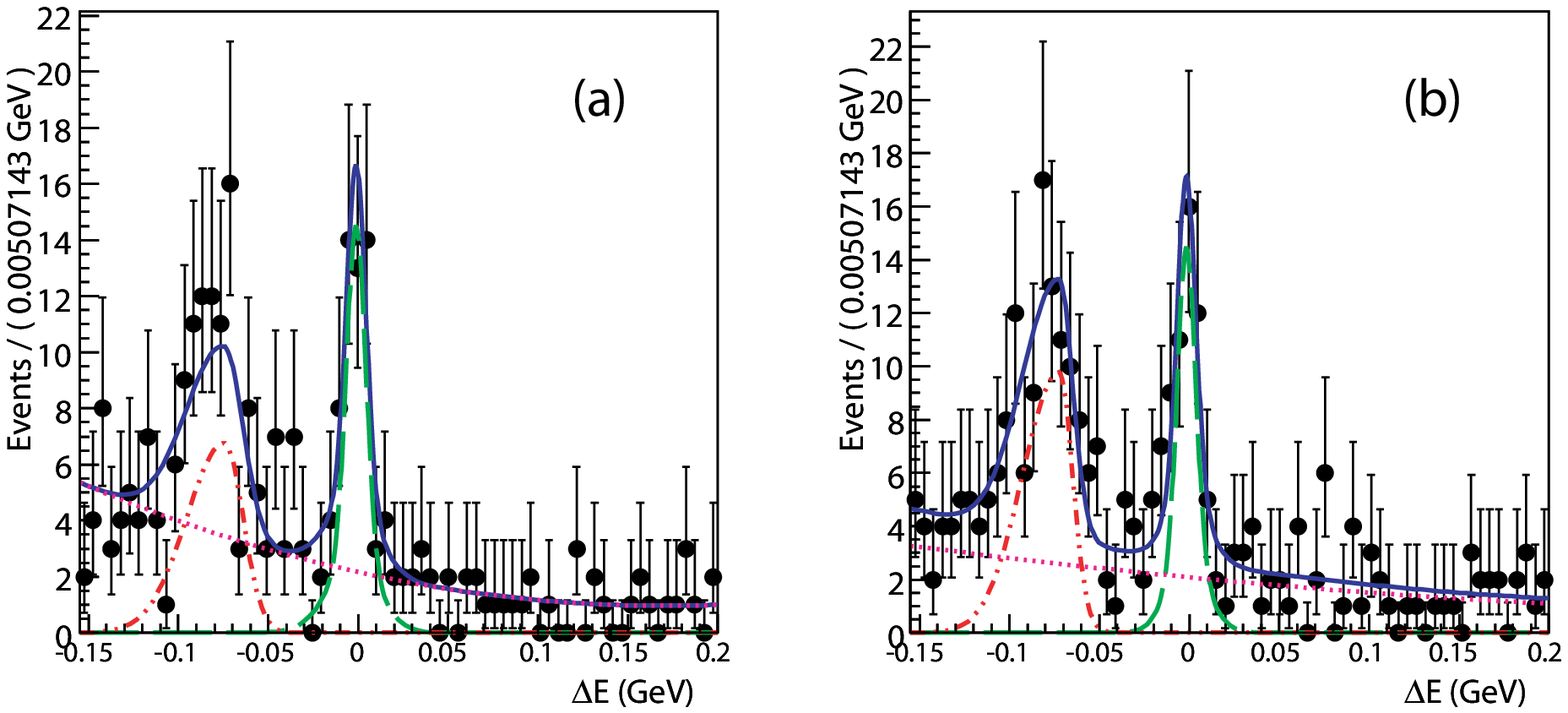}
\includegraphics[width= 0.49\textwidth]{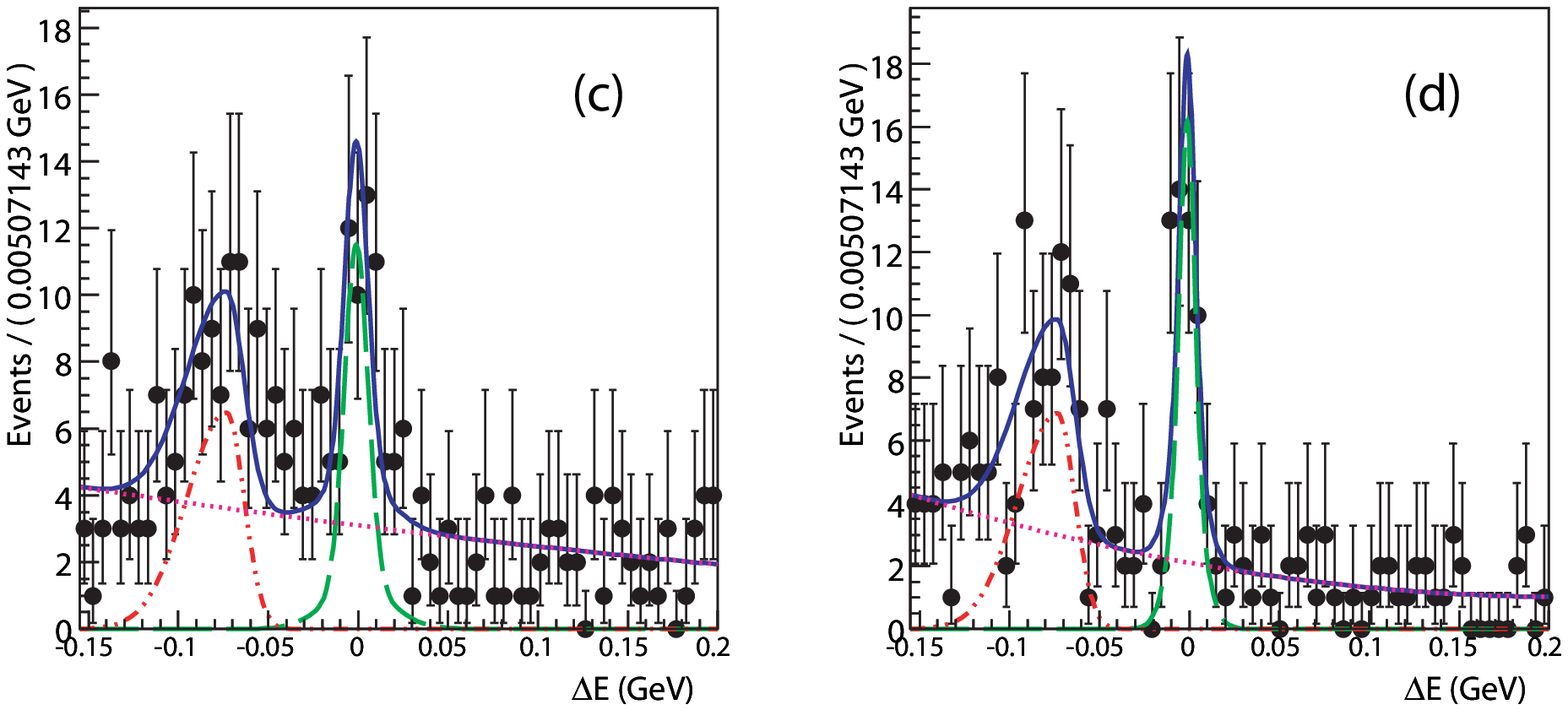}
\caption{$\Delta E$ distributions of the $B^- \to \psi(2S) \pi^-$
  candidates, reconstructed in (a) $\psi(2S) \to J/\psi (e^+e^-) \pi^+
  \pi^-$, (b) $\psi(2S) \to J/\psi (\mu^+\mu^-)\pi^+
  \pi^-$, (c) $\psi(2S)\to e^+ e^-$ and (d) $\psi(2S) \to \mu^+ \mu^-$.
  The curves show the signal (dashed) and background components
  (dot-dashed and dotted) as well as the overall fit (solid). }
\label{fig:psi}
\end{figure}
%%%%%%%%%%%%%%%%%%%%%%%%%%%%%%%%%%%%%%%%%%%%%%%%%%%%%%%%%%%%%%%%%%%%%%%%%%%%%%%%%%
The $\psi(2S)$ meson is reconstructed through the $\ell^+ \ell^-\,$ and $J/\psi \pi^+ \pi^-$ decay channels, where the
$J/\psi$ decays to $\ell^+ \ell^-$ ($\ell = e \,{\rm or}\,\mu$). 
Inclusion of charge-conjugate modes is implied throughout the paper.
Contamination from the $B^- \to \psi(2S) K^-$ decays, where a kaon is
misinterpreted as a pion, results in a peak at $\Delta E \approx
-0.07\,\rm{GeV}$, which is modeled using a Monte Carlo (MC) sample.   A fit
performed simultaneously to all the considered $\psi(2S)$ decay
modes (Fig.~\ref{fig:psi}) provides the branching fraction:
\begin{eqnarray}
\mathcal{B}(B^- \to \psi(2S) \pi^-) = 2.44 \pm 0.22 \,\rm{(stat)} \pm 0.20\,\rm{(syst)}.
\end{eqnarray}
\noindent Branching fractions for the $B^+$ and $B^-$ decays are extracted to
measure the charge asymmetry $\mathcal{A}$. Signal yields of $89 \pm
11$ and $93 \pm 11$ events  for $B^+$
and $B^-$, respectively, result in a charge asymmetry of
\begin{eqnarray}
\mathcal{A} = \frac{\mathcal{B}(B^- \to \psi(2S) \pi^-) - \mathcal{B}(B^+ \to \psi(2S) \pi^+)}{\mathcal{B}(B^- \to \psi(2S) \pi^-) + \mathcal{B}(B^+ \to \psi(2S) \pi^+)}= 0.022 \pm 0.085\,\rm{(stat)} \pm 0.016 \,\rm{(syst)},
\end{eqnarray}
which is consistent with no direct $CP$ violation.  Finally we measure
\begin{eqnarray} 
\frac{\mathcal{B}(B^- \to \psi(2S) \pi^-)}{\mathcal{B}( B^- \to \psi(2S)
  K^-)} = (3.99 \pm 0.36\,\rm{(stat)} \pm 0.17 \,\rm{(syst)}) \%,
\end{eqnarray} 
which is consistent with the theoretical prediction of the
factorization hypothesis.

\section{\boldmath{$B^0 \to D^{*+} D^{*-}$}}
\subsection{Theoretical Motivation}

The time-dependent decay
rate of a neutral $B$ meson to a $CP$ eigenstate, such as $D^{*+} D^{*-}$, is given by:
\begin{eqnarray}  
\mathcal{P}(\Delta t) &=& \frac{e^{-|\Delta t|/\tau_{B^0}}}{4
  \tau_{B^0}} \Big \{1 + q \Big [\mathcal{S} \sin(\Delta m_d \Delta t) +  \mathcal{A} \cos(\Delta m_d \Delta t) \Big ] \Big \}, 
\label{eq:basic}
\end{eqnarray}  
where $q = +1 (-1)$ when the other $B$
meson in the event decays as a $B^0$ ($\overline{B}^0$) and $\Delta t =
t_{CP} - t_{{\rm tag}}$ is the proper time difference between the two
$B$ decays in the event,  
$\tau_{B^0}$ is the neutral $B$ lifetime, $\Delta m_d$ the mass
difference between the two $B^0$ mass eigenstates. The $CP$-violating
parameters are defined as
\begin{eqnarray}  
\mathcal{S} = \frac{2 \Im(\lambda)}{|\lambda|^2 + 1}, \quad \mathcal{A}
= \frac{|\lambda|^2 - 1}{|\lambda|^2 + 1}, 
\end{eqnarray} 
where $\lambda$ is a complex observable depending on the $B^0$
  and $\overline{B}^0$ decay amplitudes to the final state and the
  relation between the $B$
  meson mass eigenstates and its flavor eigenstates.  When ignoring penguin corrections,
the SM predictions for the $CP$ parameters are $\mathcal{A}_{D^{*+}
  D^{*-}} = 0$ and $\mathcal{S}_{D^{*+} D^{*-}} = -\eta_{D^{*+} D^{*-}} \sin 2
\phi_1$, where $\phi_1 = \arg[-V_{cd}V^*_{cb}]/[V_{td}V^*_{tb}]$ and
$\eta_{D^{*+} D^{*-}}$ is the $CP$ eigenvalue of  $D^{*+} D^{*-}$, which
is $+1$ when the decay proceeds through the $S$ or $D$ wave,
or $-1$  for the $P$ wave. A significant shift of the $CP$ parameters from the SM predictions
can be a sign for New Physics.

\subsection{Yield and angular analysis}
\begin{figure}
\includegraphics[width= 0.279\textwidth]{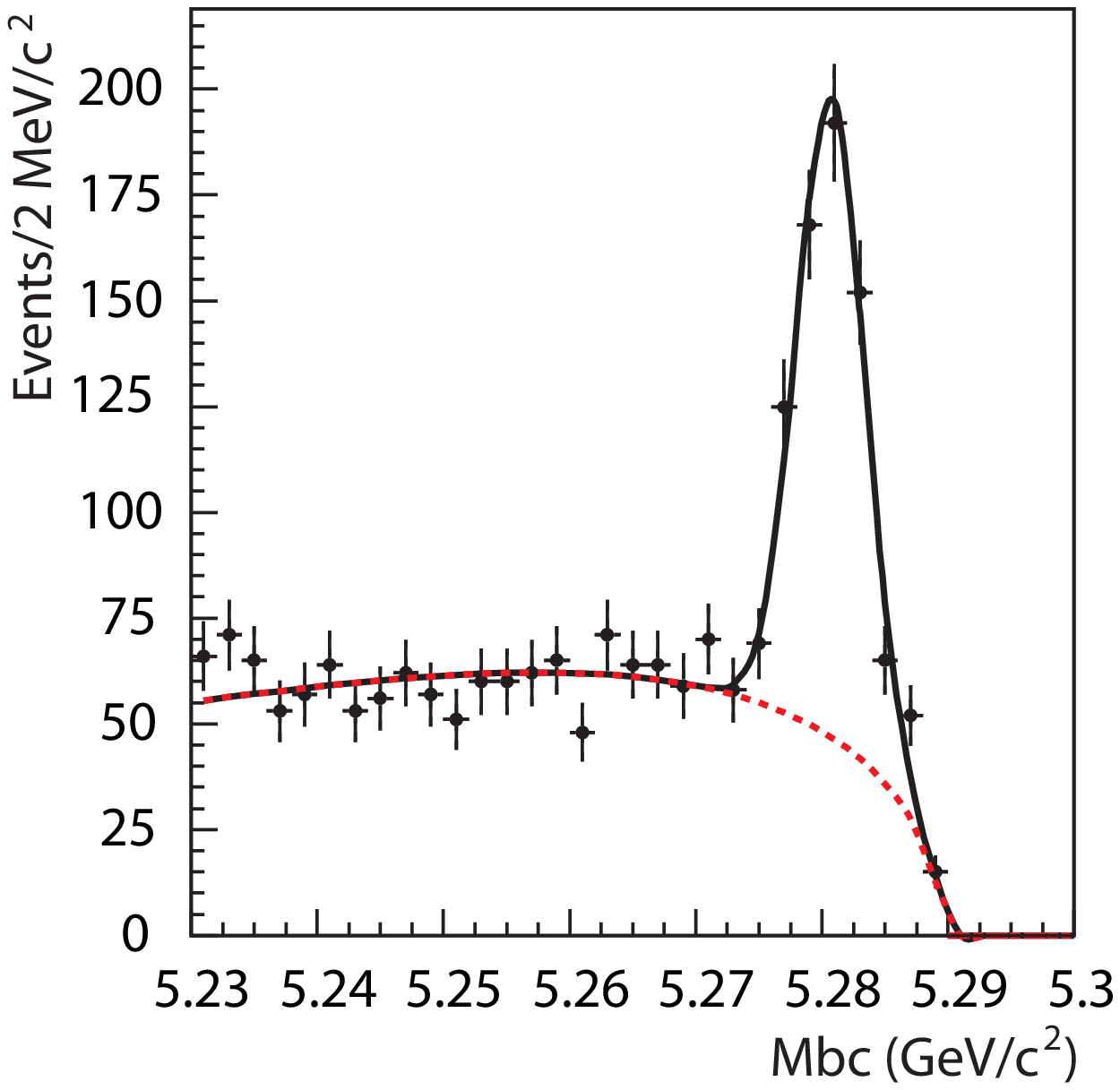}
\includegraphics[width= 0.273\textwidth]{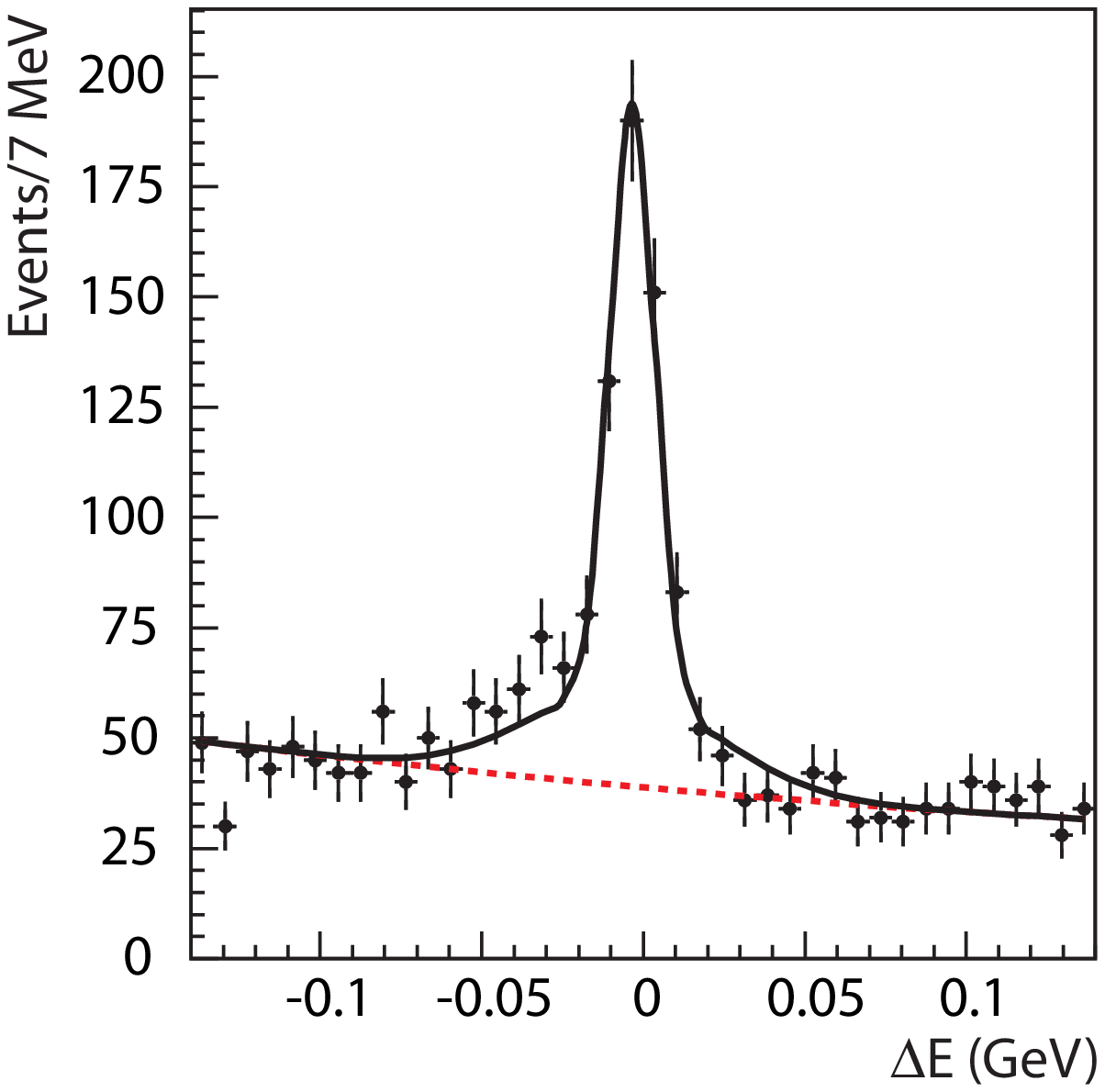}
\includegraphics[width= 0.263\textwidth]{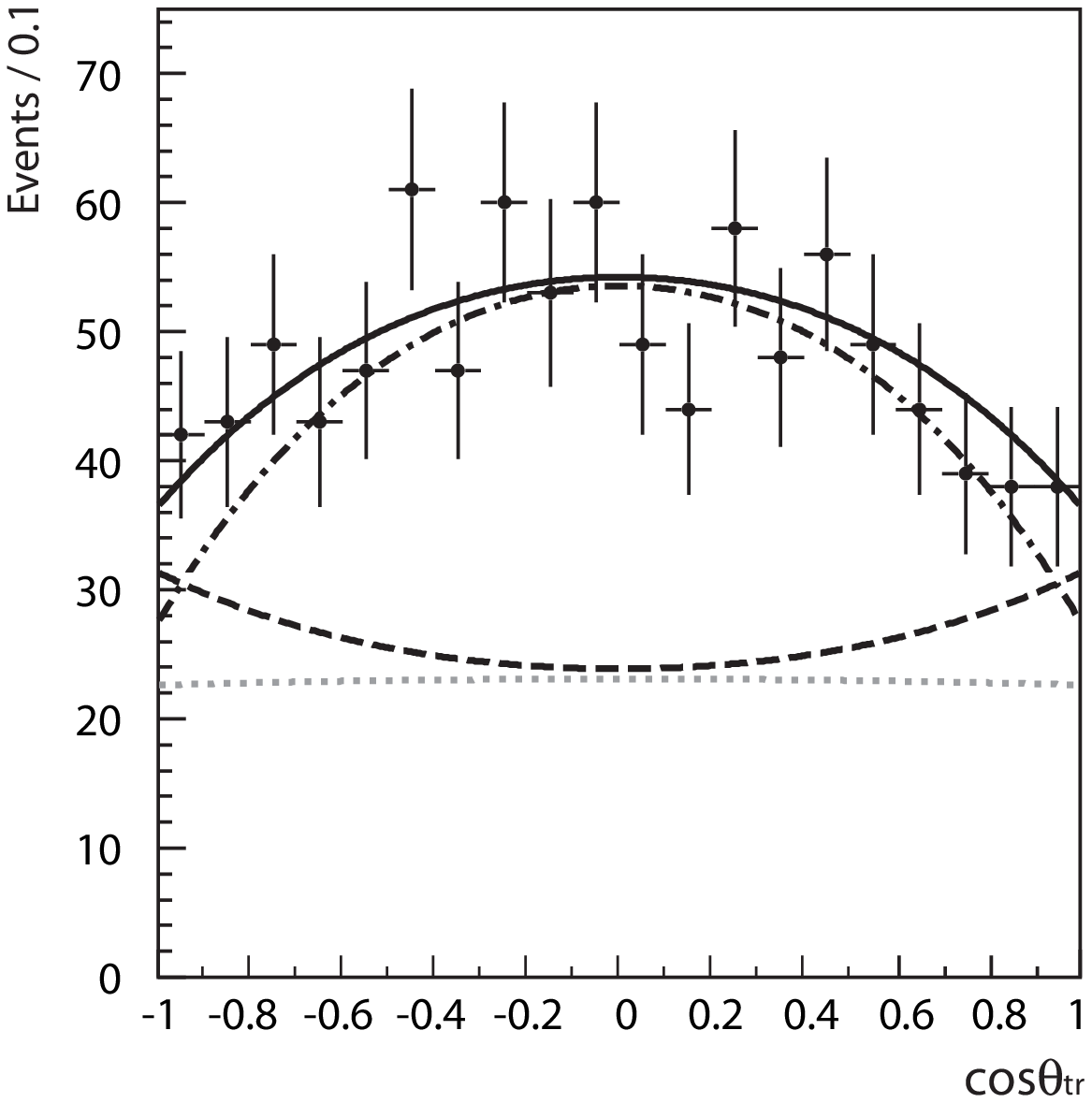}
\caption{(Left) $M_{{\rm bc}}$ distribution in the $\Delta E$ signal region. 
(Middle) $\Delta E$ distribution in the $M_{{\rm bc}}$ signal
region. (Right) $\cos\theta_{{\rm tr}}$ distribution in the signal region.
  The solid line represents the total fitted function and 
the dotted line shows the background
  contribution. In the right plot, the $CP$-even and $CP$-odd contributions are
  indicated above the background with the dashed and dot-dashed lines respectively.}
\label{fig:signal}
\end{figure}

The $D^{*\pm}$ mesons are reconstructed  in the $D^0 \pi^+$
and $D^+ \pi^0$ modes. 
%and $D^0$ mesons are reconstructed in the  $ \Km \pip$, $\Km \pip \pio$, $\Km \pip \pip \pim$, $\Ks
%\pip \pim$, $\Ks \pip \pim \pio$ and $\Kp \Km$ modes, while $\Dp$ decays are reconstructed in the $\Km \pip \pip$,  $\Ks \pip$, $\Ks \pip \pio$ and $\Kp \Km
%\pip$ modes. 
The signal is extracted from a two-dimensional
unbinned maximum likelihood fit in the $M_{{\rm bc}}$ vs. $\Delta E$
plane.  We obtain $553 \pm 30$ signal events with a 
signal purity of $55\%$.  The first two plots in Fig.~\ref{fig:signal} show the projections of the fitted $M_{{\rm bc}}$ and $\Delta E$
distributions in the signal region.

%%%%%%%%%%%%%%%%%%%%%%%%%%%%%%%%%%%%%%%%%%%%%%%%%%%%%%%%%%%%%%%%%%%%%%%%%%%%%%%%%%
%%%%%%%%%%%%%%%%%%%%%%%%%%%%%%%%%%%%%%%%%%%%%%%%%%%%%%%%%%%%%%%%%%%%%%%%%%%%%%%%%%
To obtain the $CP$-odd fraction we perform a
time-integrated angular analysis in the transversity
basis~\cite{transversity2}.  The differential decay rate as a
function of the transversity angle $\theta_{\rm{tr}}$ reads
\begin{eqnarray}  
\frac{1}{\Gamma}\frac{d\Gamma_{B \to D^{*+}
  D^{*-}}}{d\cos\theta_{{\rm tr}}} &=&
   \frac{3}{4} R_0\sin^2\theta_{{\rm tr}} 
 + \frac{3}{2} R_{\perp}\cos^2\theta_{{\rm tr}} + \frac{3}{4} R_{\parallel}\sin^2\theta_{{\rm tr}} 
%&=&   \frac{3}{4}R_{{\rm even}}\sin^2\theta_{{\rm tr}} + \frac{3}{2}
%  R_{\perp}\cos^2\theta_{{\rm tr}} \nonumber
\label{eq:polthtr}
\end{eqnarray} 
where $R_{0,\parallel}$ and $R_{\perp}$ are the $CP$-even and $CP$-odd
fractions of the three transversity components respectively.  
A one-dimensional fit of the $\cos \theta_{{\rm tr}}$ distribution
allows the extraction of the $CP$-odd fraction.  Its distortion
due to the angular resolution and the slow pion reconstruction
efficiency is modeled using signal MC samples.  The fraction $R_0/(R_{0} + R_{\parallel})$ is taken from the previous
Belle analysis~\cite{bib:miyake}.  The signal-to-background ratio is determined on an event-by-event basis 
using the $M_{{\rm bc}}-\Delta E$ distribution.  The result (shown in
the right plot of Fig.~\ref{fig:signal}) is
\begin{eqnarray}  
R_{\perp} = 0.125 \pm 0.043 {\rm (stat)} \pm
 0.023{\rm (syst)}.
\end{eqnarray}

\subsection{Time-dependent \boldmath{$CP$} violation measurement}

\begin{figure}
\includegraphics[width= 0.34\textwidth]{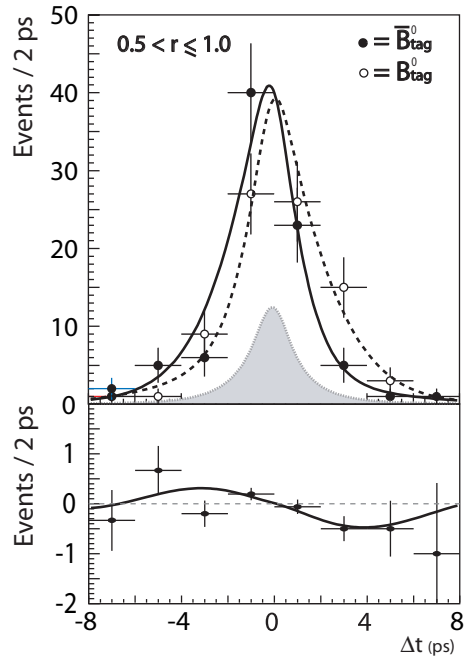}
\caption{Top: $\Delta t$ distribution of well-tagged $B^0 \to \Dstp
  \Dstm$ candidates  ($r > 0.5$) for $q = +1$ and $q = -1$.  The gray area is the background
  contribution while the solid and dashed curves are the superposition of the
  total PDFs for well-tagged $q=-1$ (solid line) and $q=+1$ (dotted
  line) events respectively.  Bottom: fitted raw asymmetry of the two top distributions.}
\label{fig:cpfit}
\end{figure}

Because the $B^0$ and $\overline{B}^0$ are
approximately at rest in the $\Upsilon(4S)$ CM frame, the $\Delta t$ value can be determined from the separation in $z$ of
the two decay vertices, $\Delta t \simeq \Delta z / (\beta \gamma c)$,
where $c$ is the speed of light.  To obtain the $\Delta t$
distribution, we reconstruct the tag-side $B$ vertex and its flavor inclusively from 
properties of particles that are not associated with the
reconstructed $B^0 \to D^{*+} D^{*-}$ decay~\cite{bib:tag}.   The tagging
information is represented by two parameters, the flavor of the
tagging $B^0$, $q$, and the tagging quality given by
seven $r$ intervals from $r = 0$ meaning no flavor discrimination to
$r = 1$ for unambiguous flavor assignment.   
Equation~\ref{eq:basic} is modified to incorporate the effect of incorrect flavor
assignment, the $CP$-odd dilution, and the description of background events.  The signal-to-background fraction is obtained on an event-by-event basis, using the previous fits of the $M_{{\rm bc}}$, $\Delta E$ and
$\cos\theta_{{\rm tr}}$ distributions.  The free parameters in the fit are
$\mathcal{A}_{\Dstp \Dstm}$ and $\mathcal{S'}_{\Dstp \Dstm} = \mathcal{S}_{\Dstp \Dstm}/\eta_{\Dstp \Dstm}$; these are determined by
maximizing an unbinned likelihood function for all events in the fit region.  The result is
\begin{eqnarray}
\mathcal{A}_{\Dstp \Dstm} &=& \,\,\,\,\,0.15 \pm 0.13 {\rm (stat)} \pm 0.04 {\rm
  (syst)}, \nonumber \\
\mathcal{S'}_{\Dstp \Dstm} &=& -0.96 \pm 0.25 {\rm (stat)}_{\,- 0.16}^{\,+0.12}
 {\rm (syst)},
\end{eqnarray}
with a statistical correlation of $10.7\%$.  The
total significance of non-zero values of $\mathcal{S}'$ and $\mathcal{A}$
 is $3.1\,\sigma$.  We define the
raw asymmetry in each $\Delta t$ bin as $(N_{+} - N_{-})/(N_{+} +
N_{-})$, where $N_{+} (N_{-})$ is the number of observed candidates
with $q = +1 (-1)$.  Figure~\ref{fig:cpfit} shows the $\Delta t$ distribution and the raw asymmetry for events with
a good-quality tag ($r > 0.5$).  Our measurement of
$\mathcal{S'}$ and $\mathcal{A}$ is consistent with the SM expectation for a
tree-dominated $b \to c \bar{c}d$ transition.
%%%%%%%%%%%%%%%%%%%%%%%%%%%%%%%%%%%%%%%%%%%%%%%%%%%%%%%%%%%%%%%%%%%%%%%%%%%%%%%%%%
%%%%%%%%%%%%%%%%%%%%%%%%%%%%%%%%%%%%%%%%%%%%%%%%%%%%%%%%%%%%%%%%%%%%%%%%%%%%%%%%%%

\section{Conclusion}
We reported a measurement of the $CP$-violating parameters in $B^- \to
\psi(2S) \pi^-$ and $B^0 \to D^{*+} D^{*-}$ decay using $657$ million $B
\overline{B}$ pairs recorded with the Belle detector.  
Both measurements are compatible with the SM predictions in absence of
penguins. The branching fraction of $B^- \to
\psi(2S) \pi^-$ is extracted as well as its ratio with respect to 
$B^- \to \psi(2S) K^-$.  The result supports the factorization hypothesis.
In the $B^0 \to D^{*+} D^{*-}$ analysis the $CP$-odd fraction is
obtained to allow for an undiluted measurement of $\sin 2 \phi_1$.

\end{document}